\documentclass[prd,twocolumn,nofootinbib]{revtex4-2}
\usepackage{preamble}
\usepackage{color}

\usepackage{aas_macros}

\begin{document}

\title{Multiband Gravitational Wave Cosmography with Dark Sirens }

\author{Brian C. Seymour\,\orcidlink{0000-0002-7865-1052}} %0000-0002-7865-1052
\email{seymour.brianc@gmail.com}
\author{Hang Yu\,\orcidlink{0000-0002-6011-6190}} %0000-0002-6011-6190
\author{Yanbei Chen\,\orcidlink{0000-0002-9730-9463}} %[0000-0002-9730-9463]
\affiliation{TAPIR, Walter Burke Institute for Theoretical Physics, California Institute of Technology, Pasadena, CA 91125, USA}

\date{\today}

%%%%%%%%%%%%%%%%
%%%%%%%%%%%%%%%%
%%% ABSTRACT %%%
%%%%%%%%%%%%%%%%
%%%%%%%%%%%%%%%%
\begin{abstract} 
    Gravitational waves might help resolve the tension between early and late Universe measurements of the Hubble constant, and this possibility can be enhanced with a gravitational wave detector in the decihertz band as we will demonstrate in this study. Such a detector is particularly suitable for the multiband observation of stellar-mass black hole binaries between space and ground, which would significantly improve the source localization accuracy thanks to a long baseline for timing triangulation, hence promoting the "dark siren" cosmology. Proposed decihertz concepts include DECIGO/B-DECIGO, TianGO, and others. We consider here the prospects of multiband observation of dark siren binaries with a variety of network configurations. We find that a multiband observation can uniquely identify a black hole binary to a single galaxy to a cosmological distance, and thus a dark siren behaves as if it had an electromagnetic counterpart. Considering only fully localized dark sirens, we use a Fisher matrix approach to estimate the error in the Hubble constant and matter density parameter. We find that a decihertz detector substantially improves our ability to measure cosmological parameters because it enables host galaxies to be identified out to a larger distance without the systematics from statistical techniques based on comparing the population distribution.
\end{abstract}
\maketitle

\maketitle

%%%%%%%%%%%%%%%%%%%%
%%%%%%%%%%%%%%%%%%%%
%%% INTRODUCTION %%%
%%%%%%%%%%%%%%%%%%%%
%%%%%%%%%%%%%%%%%%%%

\section{Introduction}\label{sec:intro}

The Hubble-Lemaître constant $H_0$ describes the current expansion rate of the universe. 
Currently, there is substantial deviation between Planck measurements of the cosmic microwave background fluctuations \cite{Planck:2018vyg} and SH0ES measurements of Type 1a supernova with the distance ladder \cite{Riess:2020fzl, Riess:2016jrr}. Notably, the Hubble tension between these early and late universe measurements differs by at least $ 4 \sigma$ \cite{Verde:2019ivm,Camarena:2021jlr}. Moreover, the tension has occurred since the first Planck results \cite{Planck:2013pxb} and strengthened with time. It is important to validate whether such Hubble tension truly exists or whether it is due to astrophysical systematics because it could signify violation from the $\Lambda$CDM concordance model \cite{Camarena:2021jlr, Vagnozzi:2019ezj, Jackson:2007ug}. One signature for departure from the concordance model would be apparent redshift evolution $H_0$ \cite{Krishnan:2020vaf}.

If the Hubble tension is proven robust with further measurements, there are a number of possible explanations with new physics. An extensive discussion of these possibilities is given in a recent review~\cite{Perivolaropoulos:2021jda}, and a comparison between many of such theories is given in~\cite{Schoneberg:2021qvd}. These proposals can be classified generally either as \textit{early-time} modification of the sound horizon or \textit{late-time} modification of the Hubble expansion.

Let us first discuss new physics before recombination which would lower the value of Hubble constant as measured by Planck. 
Early dark energy adds an additional scalar field which acts like a cosmological constant and ends after recombination \cite{Poulin:2018cxd, Agrawal:2019lmo}.
The time of matter-radiation equality can be shifted by adding additional relativistic degrees of freedom with dark radiation \cite{Buen-Abad:2017gxg, Lancaster:2017ksf, Green:2019glg} or neutrino self-interactions \cite{Cyr-Racine:2013jua, Kreisch:2019yzn}.
Finally, there are proposals that the photon mass cannot be perfectly measured due to the lifetime of the universe from Heisenberg's uncertainty principle, and this translates to uncertainty on the Hubble constant \cite{Capozziello:2020nyq, Spallicci:2021kye}.

There are also a number of ways to create a smooth late time deformation in $H(z)$ with unchanged CMB physics. 
These include phantom dark energy \cite{Alestas:2020mvb, DiValentino:2016hlg}, running vacuum model \cite{Banerjee:2019kgu, Basilakos:2015yoa}, phenomenologically emergent dark energy \cite{Li:2019yem}, vacuum phase transition \cite{DiValentino:2021izs, Caldwell:2005xb}, and a phase transition in dark energy  \cite{Banihashemi:2018has, Banihashemi:2018oxo, Benevento:2020fev}. Many of these modify the equation of state parameter of the dark energy or change how $\Omega_\Lambda(z)$ evolves with redshift.
Another way to change the evolution of $H(z)$ is by introducing additional interactions. These class includes well-known beyond GR theories such as Brans-Dicke gravity \cite{SolaPeracaula:2019zsl}, $f(R)$ gravity \cite{DAgostino:2020dhv}, and Galileon gravity \cite{Frusciante:2019puu}. Additionally, this includes interacting dark energy \cite{Wang:2004cp, DiValentino:2017iww, Yang:2019uog, Aljaf:2020eqh} where dark energy and dark matter interact, and decaying dark matter \cite{Ichiki:2004vi,Bjaelde:2012wi, Berezhiani:2015yta} where dark matter decays into an unknown dark radiation.
Finally, the homogeneous and isotropic assumption of $\Lambda$CDM can be broken with chameleon dark energy \cite{Khoury:2003rn, Khoury:2003aq, Cai:2021wgv}, cosmic voids \cite{Lombriser:2019ahl}, and inhomogeneous causal horizons \cite{Fosalba:2020gls}.

The detection of gravitational waves (GW) can provide an independent \textit{late} universe measurement of the Hubble constant.
By measuring the expansion rate in the late universe, GW could be used as an independent measurement of the Hubble constant from SH0ES. Furthermore, with a distribution of GW events at low redshifts ($z \sim 0.0 - 0.5$), an anomalous evolution of the expansion rate could be observed.
In particular, the luminosity distance of the source can be obtained from the measured gravitational waveform \cite{Schutz:1986gp}. 
A Hubble constant measurement can be readily attained from a standard siren: a binary neutron star (BNS) merger with a coincident EM counterpart \cite{Schutz:1986gp, Holz:2005df}. With the optical measurement of the redshift from EM followup and luminosity distance measurement from the GW detector, one can directly measure the Hubble constant. Indeed, the Hubble constant was measured with the BNS GW170817 \cite{LIGOScientific:2017adf} and its corresponding EM counterpart \cite{Nicholl:2017ahq,Coulter:2017wya,LIGOScientific:2017vwq}. 

However, only a small number of GW events are expected to be bright BNS mergers with EM counterparts.
The majority of observed GW events are binary black hole (BBH) events without EM counterparts, which are thus known as {\it dark sirens}.
Notably, many BBH events have already been detected and cataloged \cite{LIGOScientific:2020ibl,LIGOScientific:2021djp, Venumadhav:2019lyq, Olsen:2022pin, Nitz:19, Nitz:20}. 
Dark sirens can measure the Hubble constant by statistical techniques using galaxy catalogues \cite{Schutz:1986gp,DelPozzo:2011vcw,Chen:2017rfc, Fishbach:19, Scelfo:20, Finke:21, CigarranDiaz:22, Mukherjee:22} and features in the mass distribution~\cite{Taylor:12, Farr:19, Mastrogiovanni:21, Ezquiaga:2020tns, Ezquiaga:2022zkx, Yu:22}. These statistical techniques can be further extended with realistic galaxy clustering which provide improvements in identifying the redshift due to galaxy density correlations \cite{MacLeod:2007jd, Nair:2018ign,Gray:2019ksv, Mukherjee:21}. These statistical techniques have been applied to the GWTC-3 catalog, and the Hubble constant is measured as $H_0 = 68^{+13}_{-12} \kmsmpc $ using only dark sirens~\cite{LIGOScientific:2021aug} at 68\% credible level. By combining the statistical method with the only standard siren GW170817, the Hubble constant is measured as $H_0 = 68^{+8}_{-6} \kmsmpc $. For reference, GW170817 alone gives a Hubble constant value of $H_0 = 69^{+17}_{-8} \kmsmpc $ \cite{LIGOScientific:2017adf}. We need to bear in mind that the statistical dark siren approach relies fundamentally on population models so there is additional systematic uncertainties \cite{LIGOScientific:2021aug, Yu:22}. In contrast, the Planck Hubble constant measurement was $H_0 = 67.4^{+0.5}_{-0.5} \kmsmpc $ \cite{Planck:2018vyg} and the SH0ES measurement was $H_0 = 72.5^{+1.0}_{-1.0} \kmsmpc $ \cite{Riess:2021jrx} which corresponds to the $ 4 \sigma$ tension~\cite{Verde:2019ivm}. 

One new potential class of detector is one in the decihertz range (0.01 - 1 Hz), and such a detector may aid in measuring the Hubble constant. This detector would lie in between the millihertz LISA band \cite{LISA:2017pwj} and the 10 - 1000 Hz ground band.
A decihertz detector has many advantages for measuring the Hubble constant. First, it would provide early warning for BNS mergers which would help guarantee EM identification \cite{Kuns:2019upi, Sedda:2019uro}. Second, a joint decihertz detection would improve the parameter estimation for stellar mass BBH by measuring their waves several years before they enter into the ground band \cite{Kuns:2019upi, Yu:2020dlm}. Since statistical approaches to dark sirens are degraded by having too many galaxies inside of the localization volume, having a better angular localization will significantly help measure the cosmological parameters. Furthermore, the fascinating possibility of a multiband detection exists where a decihertz detector observes a BBH inspiral and then the ground based detectors measure the merger and ringdown. A decihertz multiband detection has been found to substantially improve parameter estimation accuracy~\cite{Kuns:2019upi}. 
By combining decihertz and ground detectors, the detector network can uniquely localize a BBH to a its host galaxy without any EM counterpart.
While a ground network can do this on its own~\cite{Chen:2016tys}, the addition of a decihertz detector will significantly increase the range at which the BBH can be localized. In this way, a multiband detection of a BBH can behave like a standard siren. 

Right now, there are a number of existing and proposed gravitational wave detectors. 
Advanced LIGO \cite{LIGOScientific:2014pky}, Advanced Virgo \cite{VIRGO:2014yos}, and KAGRA \cite{KAGRA:2018plz} are operating ground based gravitational wave detectors and are second-generation (2G) detectors.
Following the 2G detectors, LIGO Voyager aims to maximize the reach of existing LIGO observatory facilities by adding cryogenic operation, heavier silicon test masses, and improved quantum squeezing~\cite{Adhikari:17, LIGO:2020xsf}.
Einstein Telescope \cite{Punturo:2010zz} and Cosmic Explorer \cite{Reitze:2019iox} are the 3rd generation of ground-based detectors with planned arm lengths of 10 km and 40 km respectively which aim to begin observation in the mid 2030s. 3G detectors can break the distance-inclination degeneracy using higher order spherical harmonic modes which would improve Hubble constant measurement \cite{Borhanian:2020vyr}.

At frequencies below $\sim 1$\,Hz, detecting gravitational waves may best be carried out in space due to technical challenges~\cite{Hall:2020dps, Harms:2013raa}.
LISA \cite{LISA:2017pwj, LISACosmologyWorkingGroup:2022jok}, TianQin \cite{TianQin:2015yph}, and Taiji \cite{Hu:2017mde, Ruan:2018tsw}, are proposed space based detectors which focus on the $\sim 10^{-3} - 10^{-1}$ Hz bands.
LISA can measure the Hubble constant with dark sirens~\cite{DelPozzo:2017kme} with accuracy of 5\% and may be able to measure it with EMRIs \cite{Laghi:2021pqk} to an accuracy of 1\% to 3\%, though it is likely that ground detectors will surpass this by the time it operational. Ref.~\cite{Zhu:2021bpp} studied measuring the Hubble constant measurement with TianQin and LISA/Einstein Telescope. In the far future, there are proposals for a microhertz GW detector \cite{Sesana:2019vho}. At very low frequencies, it may be possible for a pulsar timing array to measure the effect of a super massive black hole binary \cite{Spallicci:2011nr}.
Furthermore, there are a number of space based plans for a decihertz detector in the $0.01 - 1$ Hz band. The Japanese detector DECIGO is an ambitious prospect that consists of three clusters of interferometers with a 1000km arm length \cite{Kawamura:2020pcg, Sato:2017dkf, Seto:2001qf}. Big Bang Observer is concept like DECIGO by the European Space Agency \cite{Crowder:2005nr}. Previous work found that Big Bang Observer alone would provide precision cosmological tests by measuring and localizing nearly every GW event in the universe \cite{Cutler:2009qv}.
Recently, Ref.~\cite{Liu:2022rvk} studied the capabilities of DECIGO and other decihertz detectors to measure the Hubble constant.
B-DECIGO is a planned pathfinder mission of DECIGO with a single interferometer and a 100 km arm length \cite{Kawamura:2020pcg, Sato:2017dkf}.
Finally, TianGO is a space based decihertz concept which is designed with nearer-term technology \cite{Kuns:2019upi,kunsthesis}.

\begin{figure}[h]
    \includegraphics[width=\linewidth]{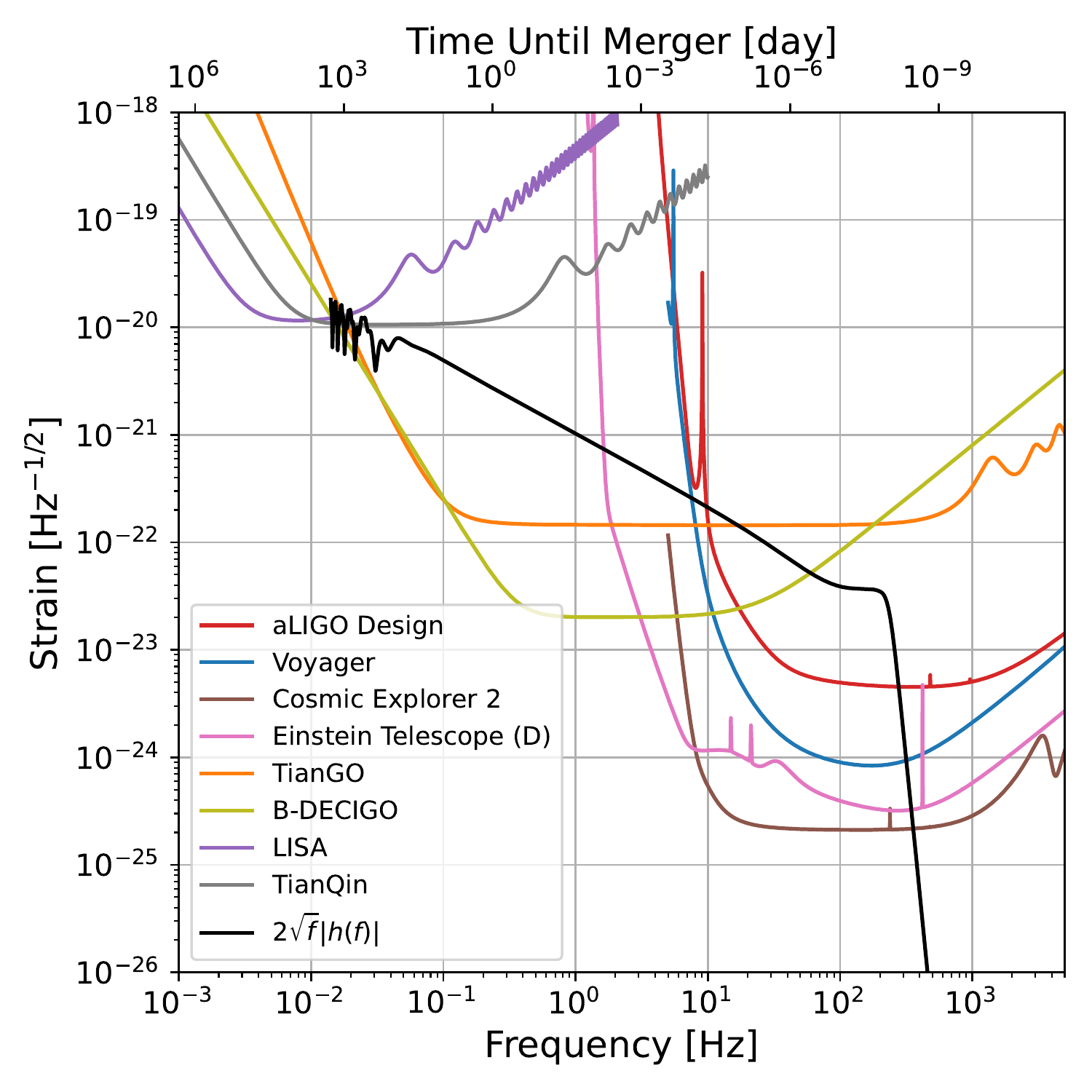}
    \caption{Comparison of detectors with a sample waveform. We plot the instrumental sensitivities for TianGO (orange), LIGO Voyager (blue), LISA (purple), aLIGO design sensitivity (red), Cosmic Explorer 2 (brown), Einstein Telescope D (pink), TianQin (gray), and B-DECIGO (yellow). We also show a sample TianGO waveform for a typical BBH merger (black) at $z = 0.3$, $\mathcal{M}_c = 25 M_\odot$, $q = 1.05$, and $T_\text{obs} = 5 \text{ yr}$ assuming observed by a TianGO-like detector. On the top axis, we give the time until merger. 
    }
    \label{fig:detector-asd-waveform}
\end{figure}

For this analysis, we study how well we can measure the expansion rate of the universe by measuring BBH with future ground detectors and decihertz concepts. We consider two representative decihertz detectors: (i) TianGO in the LIGO Voyager era, and (ii) B-DECIGO in the ET/CE era. TianGO is chosen because it represents a possible near term decihertz detector. In such a timescale, it would be operational in late 2020s/early 2030s and be working with the LIGO Voyager network. B-DECIGO is a longer term prospect, which would be operational in the late 2030s.

We forecast how well a dark siren can be localized with the Fisher matrix formalism~\cite{Finn:1992wt, Cutler:1994ys} with both detector setups. If such a comoving volume contains only one galaxy, we consider the dark siren to be \textit{localized}. We consider the case where localized events will have measured redshift due to either spectroscopic follow-up or from a complete galaxy catalog. We find that adding a decihertz detector to the network improves the range at which a dark siren can be localized. We then constrain the Hubble constant and matter density parameter by stacking the localized dark siren events together with the BBH merger rate inferred by LIGO/Virgo/KAGRA~\cite{LIGOScientific:2021psn}. We assume that the Hubble constant and matter density are the Planck values and fix all other cosmological parameters. Our study motivates how a decihertz detector can complement the cosmological measurement capabilities of ground based detectors. 

The rest of the paper is organized as follows. In Sec.~\ref{sec:BBH-waveform}, we describe the observed strain in a space based detector, and we use the Fisher matrix formalism to forecast the measurement uncertainties with a multiband detection. In Sec.~\ref{sec:cosmo-constraints}, we describe how we stack localized events together and the forecast dark siren constraints on the Hubble constant and matter density parameter for various detector setups. We then conclude this work in Sec.~\ref{sec:conclusion}. Finally, App.~\ref{app:antenna} delves into the space-based waveform specifics, and App.~\ref{app:bayes} justifies the conservative approach of considering only localized dark sirens.
Throughout the work, we use $G=c=1$.

%%%%%%%%%%%%%%%%%%%%
%%%%%%%%%%%%%%%%%%%%
%%% Section 2 %%%
%%%%%%%%%%%%%%%%%%%%
%%%%%%%%%%%%%%%%%%%%
\section{Measurement of a Binary Black Hole}\label{sec:BBH-waveform}

\subsection{TianGO Waveform}

Let us first model the waveform in a space detector. TianGO is orbiting the sun at an inclination of $60^\circ$, similar to the orbit of LISA~\cite{Dhurandhar:05}. Thus, there are two coordinate frames for the geometry of TianGO. We denote the ecliptic frame to have basis $(\ubarvect x, \ubarvect y, \ubarvect z)$ where $\ubarvect z$ is normal to the orbit of the earth. The frame with $(\uvect x, \uvect y, \uvect z)$ is fixed on the center of TianGO with $(\uvect x, \uvect y)$ oriented along its two arms. We denote $\uvect N$ as the line of sight vector and $\uvect L$ is the direction of binary angular momentum.
We can write the waveform as~\cite{Yu:2020dlm}
\begin{equation}\label{eq:space-waveform}
    \tilde h(f) = \Lambda(f) e^{- i \left[\Phi_P(f) +\Phi_D(f)\right]} \tilde h_c(f) \,,
\end{equation}
where $\tilde h_c(f)$ is the carrier waveform, $\Lambda(f)$ is the amplitude in Eq.~\eqref{eq:lambdaf}, $\Phi_P(f)$ is the polarization phase in Eq.~\eqref{eq:phip}, and $\Phi_D(f)$ is the phase modulation due to Doppler effect in Eq.~\eqref{eq:phid}.
The carrier waveform is independent of the antenna patterns and only depends on the intrinsic parameters $(\mathcal{M}_z, q, D_L, t_c, \phi_c)$ where $\mathcal{M}_z = (1 +z) \mathcal{M}_c$ is the detector frame chirp mass, $q$ is the mass ratio, $D_L$ is the luminosity distance, and $t_c, \phi_c$ are the time and phase of coalescence. Because we wish to model the gravitational waveform over the frequencies in both TianGO and Voyager, the carrier waveform is modeled with a phenomenological waveform that combines inspiral, merger and ringdown. Specifically, we use a \verb!IMRPhenomD! waveform \cite{Khan:2015jqa, Husa:2015iqa}.

The notable difference for a space-based detector compared to a ground one is that the orientation and location change with time. Thus, the amplitude and polarization phase which characterize the antenna patterns acquire a frequency dependence and are derived in \cite{Cutler:1997ta, Apostolatos:1994mx} for a space based detector. We write them as
\begin{align}
    \Lambda(f) &= \left[ A_+^2 F_+^2(f) + A_\times^2 F_\times^2(f) \right]^{1/2} \, , \label{eq:lambdaf}\\
    \Phi_P &= \arctan \left[ \frac{- A_\times F_\times(f)}{A_+ F_+(f)} \right] \, \label{eq:phip}.
\end{align}
$F_{+,\times}(\phi_S, \theta_S, \psi_S)$ are the detector beam pattern coefficient where $(\phi_S, \theta_S)$ are the direction of $\uvect N$ in the TianGO corotating frame and the barred ones denote quantities in the ecliptic frame, and $\psi_S$ is the polarization phase. The polarization amplitudes are $A_+ = 1 + (\uvect L \cdot \uvect N )^2$ and $A_\times = 2 \uvect L \cdot \uvect N$. 
Additionally, there is a phase modulation due to the Doppler effect induced by the orbital motion of the detector (which we have assumed to be a heliocentric one), 
\begin{align}
    \Phi_D(f) &= 2 \pi f \tau \, \label{eq:phid},\\
    &= 2 \pi f R_\text{AU} \sin \bar\theta_S \cos \left( \bar\phi_t(f) - \bar\phi_S \right) \, ,
\end{align}
where $\tau = - \vect d \cdot \uvect N$, $\vect d$ is the vector from barycenter to detector, $R_{\rm AU}$ is one AU, and $\bar\phi_t(f)$ is the azimuthal location of the solar orbit of the detector. The explicit expressions for $F_{+,\times}$, $\uvect L \cdot \uvect N$, $\bar\phi_t(f)$ are given in App.~\ref{app:antenna}.
The ground waveforms are the same as Eq.~\eqref{eq:space-waveform}, but they are approximated as $f\rightarrow \infty$ for $\Lambda(f),\Phi_P(f),\Phi_D(f)$ since the antenna patterns are nearly constant while it is in band.

In Fig.~\ref{fig:detector-asd-waveform}, we plot a sample TianGO BBH waveform, along with the sensitivity of some gravitational wave detectors. This waveform terminates on the left side because of the 5 year observation time. It exhibits amplitude modulation around $f\sim 2 \cdot 10^{-2}$ Hz because TianGO's orientation $\uvect N$ is changing with a period of a year.

%%%%%%%%%%%%%%%%%%%%
%%%%%%%%% PE %%%%%%%
%%%%%%%%%%%%%%%%%%%%
\subsection{Parameter Estimation Background}

Let us now describe how we use the Fisher analysis to estimate parameter uncertainties. The Fisher matrix formalism provides a useful approximation to parameter estimation in the high SNR limit \cite{Finn:1992wt,Cutler:1994ys, Vallisneri:2007ev}. We consider a binary with parameters $\vect{\theta}^a$ and 
\begin{equation}
    \vect{\theta}^a = \left( \ln \mathcal{M}_z, q, \ln D_L, t_c, \phi_c, \bar\phi_S,\bar\theta_S,\bar\phi_L,\bar\theta_L \right) \, .
\end{equation}
The variance for a specific parameter $\vect{\theta}^a$ is found on the diagonal of the inverse of the Fisher matrix
\begin{equation}
    \Delta \vect{\theta}^a = \sqrt{\left( \Gamma^{-1} \right)_{aa}} \, ,
\end{equation}
where the Fisher information matrix is defined as
\begin{equation}
    \Gamma_{ab} \equiv \left( \frac{\partial \tilde h}{\partial \vect{\theta}_a} \Big| \frac{\partial \tilde h}{\partial \vect{\theta}_b}\right) \, ,
\end{equation}
and the waveform template $\tilde h(f,\vect\theta)$ is a function of frequency $f$ and parameters $\vect \theta$. The inner product between two signals $\tilde h(f), \tilde g(f)$ is defined as
\begin{equation}
    \left( \tilde g \big|\tilde h \right) = 4 \, \text{Re} \int_0^\infty \frac{\tilde g^{\ast}(f) \tilde h(f)}{S_n(f)} df
\end{equation}
where $S_n(f)$ is the detector noise spectral density. In the case of a network of detectors, we sum the individual Fisher matrix for each detector $d$
\begin{equation}
    \left( \Gamma_{ab} \right)^\text{net} = \sum_d \Gamma_{ab}^d \,.
\end{equation}

\subsection{Results from Parameter Estimation}

To understand how a decihertz detector can enhance the parameter estimation of a BBH, we examine the results obtained using TianGO with the HLI Voyager network. The luminosity distance is defined by
\begin{equation}
    D_L(z) = \frac{1+z}{H_0}\int_0^z \frac{dz'}{E(z')}
\end{equation}
where 
\begin{equation}
    E(z) \equiv \sqrt{\Omega_m \left( 1+z \right)^3 + \Omega_\Lambda } \, .
\end{equation}
For precision tests of cosmology, we are mostly interested in the luminosity distance accuracy and volume localization.
The size of the solid angle ellipse $\Delta \Omega$ can be expressed by \cite{Cutler:1997ta}
\begin{equation}
    \Delta \Omega = 2 \pi \sin \bar\theta_S \sqrt{ \Sigma_{\bar\phi_S\bar\phi_S} \Sigma_{\bar\theta_S\bar\theta_S} - \left( \Sigma_{\bar\theta_S\bar\theta_S} \right)^2} \, .
\end{equation}
The uncertainty in comoving volume can be related to the angular uncertainty by Eq.~(28) of Ref.~\cite{Hogg:1999ad}
\begin{equation}
    \Delta V_{\mathrm{C}}=\frac{D_{L}^{2}}{(1+z)^{2}} \Delta \Omega \Delta D_C \, ,
\end{equation}
where the comoving distance equals $D_C = D_L / \left( 1+z \right)$. Using a change of variables, the comoving volume uncertainty can be rewritten as 
\begin{equation} \label{eq:Vc}
    \Delta V_{\mathrm{C}}=\frac{D_L^2}{(1+z)^{3} + D_{L} H(z) \left( 1+z \right)} \Delta \Omega \Delta D_{L} \, ,
\end{equation}
where $H(z) = H_0 E(z)$.

Systematic errors beyond the detector sensitivity can degrade the accuracy of the luminosity distance. The first of which is the gravitational lensing which changes the luminosity distance. We use the fit from \cite{Hirata:2010ba}
\begin{equation}
    \frac{\left( \Delta D_L \right)_\text{lens}}{D_L}=0.066\left[\frac{1-(1+z)^{-0.25}}{0.25}\right]^{1.8} \, .
\end{equation}
Once a particular galaxy is identified, the peculiar velocity adds uncertainty to the amount of cosmological redshift. The measured redshift is the sum of the cosmological and Doppler redshift. We can express the peculiar velocity systematic error as \cite{Gordon:2007zw}
\begin{equation}
    \frac{\left( \Delta D_L \right)_\text{pv}}{D_L} = \Big|1 - \frac{\left( 1+z \right)^2 }{D_L H(z)} \Big| \sigma_v \, ,
\end{equation}
where we have assumed $\sigma_v = 200 \text{ km s}^{-1} / c$. The relative magnitude of this effect decreases rapidly with distance since the cosmological redshift increases while the RMS peculiar velocity is approximately constant.

\begin{figure}[h]
    \centering
    \includegraphics[width=0.8\linewidth]{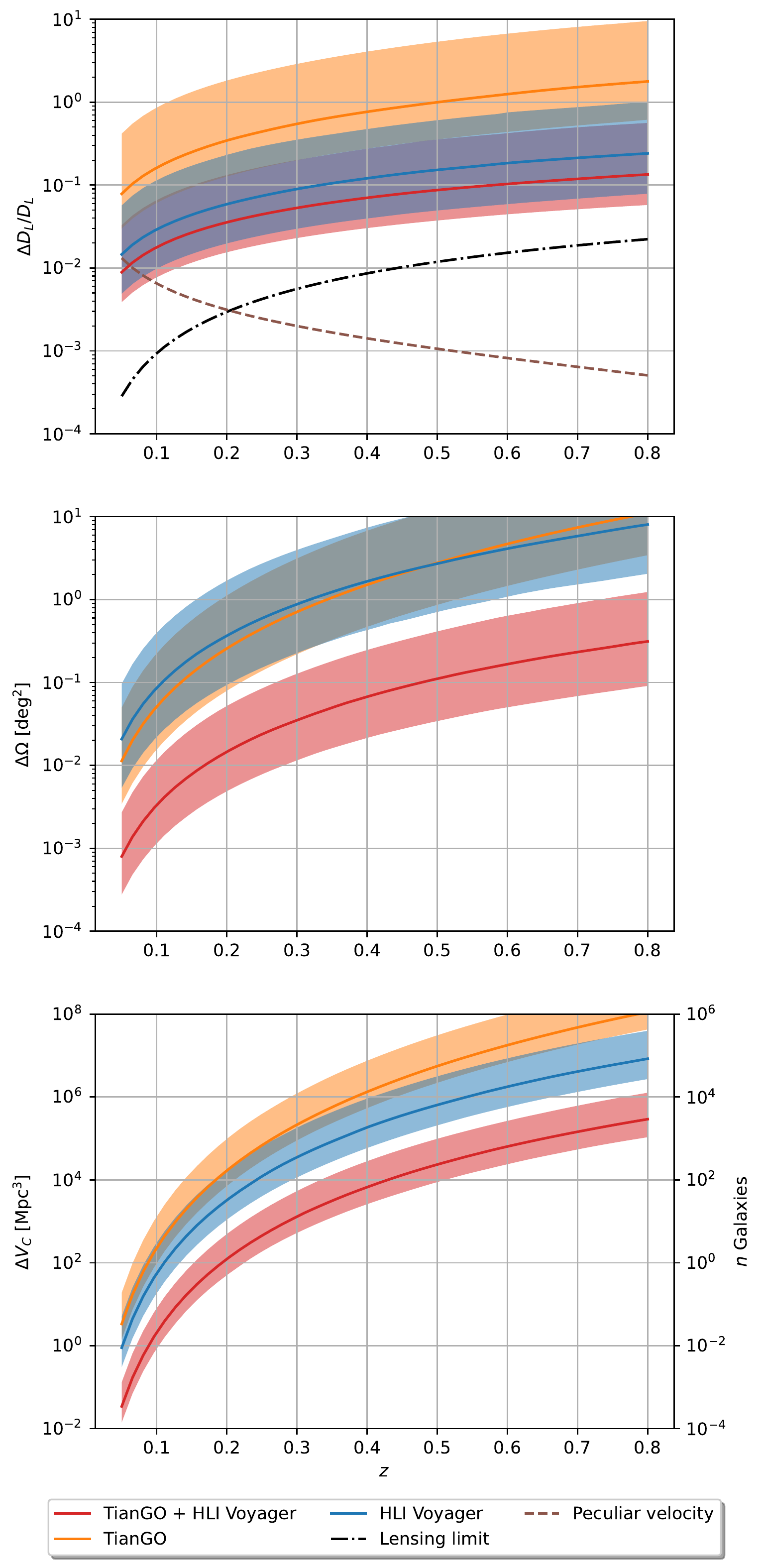}
    \caption{\label{fig:PE}
        Measurement accuracy for luminosity distance, angular resolution and comoving volume localization versus redshift. We plot these measurement uncertainties for TianGO + HLI Voyager (red), HLI Voyager (blue), and TianGO (orange). Because we randomize over the angular extrinsic parameters $(\bar\phi_S,\bar\theta_S, \bar\phi_L, \bar\theta_L)$, we plot both the median measurement with the line and the shaded region where $80\%$ of binaries lie.
        We use $\mathcal{M} = 25 M_\odot$, $q = 1.05$, $t_a = 5^\circ$ and $T_\text{obs} = 5 \text{ yr}$. 
        We use a galaxy number density per comoving volume of $n_\text{gal} = 0.01 \text{ gal}/\text{Mpc}^3$ to convert comoving volume localization to estimate our ability to identify the GW source.
        }
\end{figure}

Figure~\ref{fig:PE} gives the measurement accuracy for luminosity distance, angular resolution, and spatial localization. We considered a binary of $\mathcal{M}_c = 25 M_\odot$, $q = 1.05$, a trailing angle between earth and TianGO of $t_a = 5^\circ$, and a $5$ year observation. 
The measurement accuracy strongly depends upon inclination $\iota$ of the binary, in addition to orientation of the detector network at merger. Therefore, we randomize over $(\bar\phi_S,\bar\theta_S, \bar\phi_L, \bar\theta_L)$ in the figure. The line represents the median measurement accuracy while the shaded region contains $80 \%$ of possible systems. 
While we use a 5 year observing time for TianGO, the TianGO's parameter estimation isn't particularly sensitive to the observing time as long as it's above $\sim 1 \text{ week}$ as most of the SNR comes from frequencies above $0.1 \text{ Hz}$ (see Fig.~\ref{fig:detector-asd-waveform}).

In the top part of Fig.~\ref{fig:PE}, we show the fractional uncertainty in the luminosity distance $\Delta D_L / D_L$ versus redshift. One can see that the addition of TianGO doesn't significantly improve the ability to measure the luminosity distance compared with the HLI network. Most of the SNR from the event comes from the ground network, so the addition of TianGO improves the luminosity distance measurement by a factor of only 1.5\footnote{Note that we have published a previous paper where we found that TianGO improved the luminosity distance measurement of the HLI Voyager network (Fig.~3 and Fig.~13 of \cite{Kuns:2019upi}). There was an error in the space waveform code.}. We also plot the lensing and peculiar velocity systematic errors here. We see that the systematic error due to peculiar velocity is only large enough to affect our measurement for very close events. Meanwhile, the effect of lensing is negligible and can be ignored in the future sections about cosmology.

In the middle panel of Fig.~\ref{fig:PE}, we give the angular resolution $\Delta \Omega$ versus redshift. We see an angular resolution improvement by a factor of 20 for the addition of TianGO to the HLI Voyager network. The long baseline between earth and TianGO is responsible for this upgraded sky localization sensitivity. 

Finally, let us describe the comoving volume localization in the bottom panel of Fig.~\ref{fig:PE}. We plot the comoving volume localization from Eq.~\eqref{eq:Vc}, and find that adding TianGO improves the comoving volume localization by a factor 30. 
We use a comoving galaxy density of $n_\text{gal} = 0.01 \text{ gal}/\text{Mpc}^3$ \cite{Chen:2016tys}. This corresponds to the number density which are about 25\% as bright as the Milky Way. This is because the majority of the GW are expected to come from galaxies at least this luminous \cite{Chen:2017rfc}.
If $n_\text{gal} \Delta V_C <1$, we say the galaxy was localized. Using this criterion, we find that HLI Voyager can localize galaxies up to $z\sim 0.15$, while TianGO + HLI Voyager can localize them up to $z\sim 0.30$. Note that error bands are large and asymmetric because the line-of-sight direction and detector configuration greatly affect the measurement accuracy. For example, a gravitational wave that is face-on to the ecliptic plane would be poorly localized by TianGO since the Doppler term does not give any information, while a GW coming edge-on will measure $\Delta\Omega$ well (and correspondingly $\Delta V_C \sim D_L^2 \Delta D_L \Delta \Omega$).

\subsection{Event Rate}

To infer cosmological parameters, we stack all dark siren events that the network can localize.
Let us now estimate how many dark sirens can be localized.
First, the merger rate density $\mathcal{R}(z)$ describes the number of mergers in a comoving volume per year. We model it with a power law model and choose with $\kappa = 2.7$ so that it corresponds to the Madau-Dickinson star formation rate \cite{Madau:2014bja} 
\begin{equation}
    \mathcal{R}(z) = \mathcal{R}_0 \left( 1 + z \right)^{\kappa} \, .
\end{equation}
Since this is the source frame merger rate density, an additional factor of $1/(1+z)$ is needed to convert time from the source frame to the detector frame. Therefore, we write the detector-frame merger rate of sources with $z<z_m$ as
\begin{equation} \label{eq:merger-rate}
    R_\text{obs}(z_m) = \int_0^{z_m} \mathcal{R}(z') \frac{1}{1+z'} \frac{d V_c}{dz'} dz' \, ,
\end{equation}
where 
\begin{equation}
    \frac{d V_c}{d z} = \frac{4 \pi }{H_0} \frac{d_c^2(z)}{E(z)} \, .
\end{equation}
We use the BBH merger rate $\mathcal{R}_0 = 20 \text{ Gpc}^{-3} \text{yr}^{-1}$ and $\kappa = 2.7$ which consistent with GWTC-3 \cite{LIGOScientific:2021psn}.

\begin{figure}
    \centering
    \includegraphics[width=\linewidth]{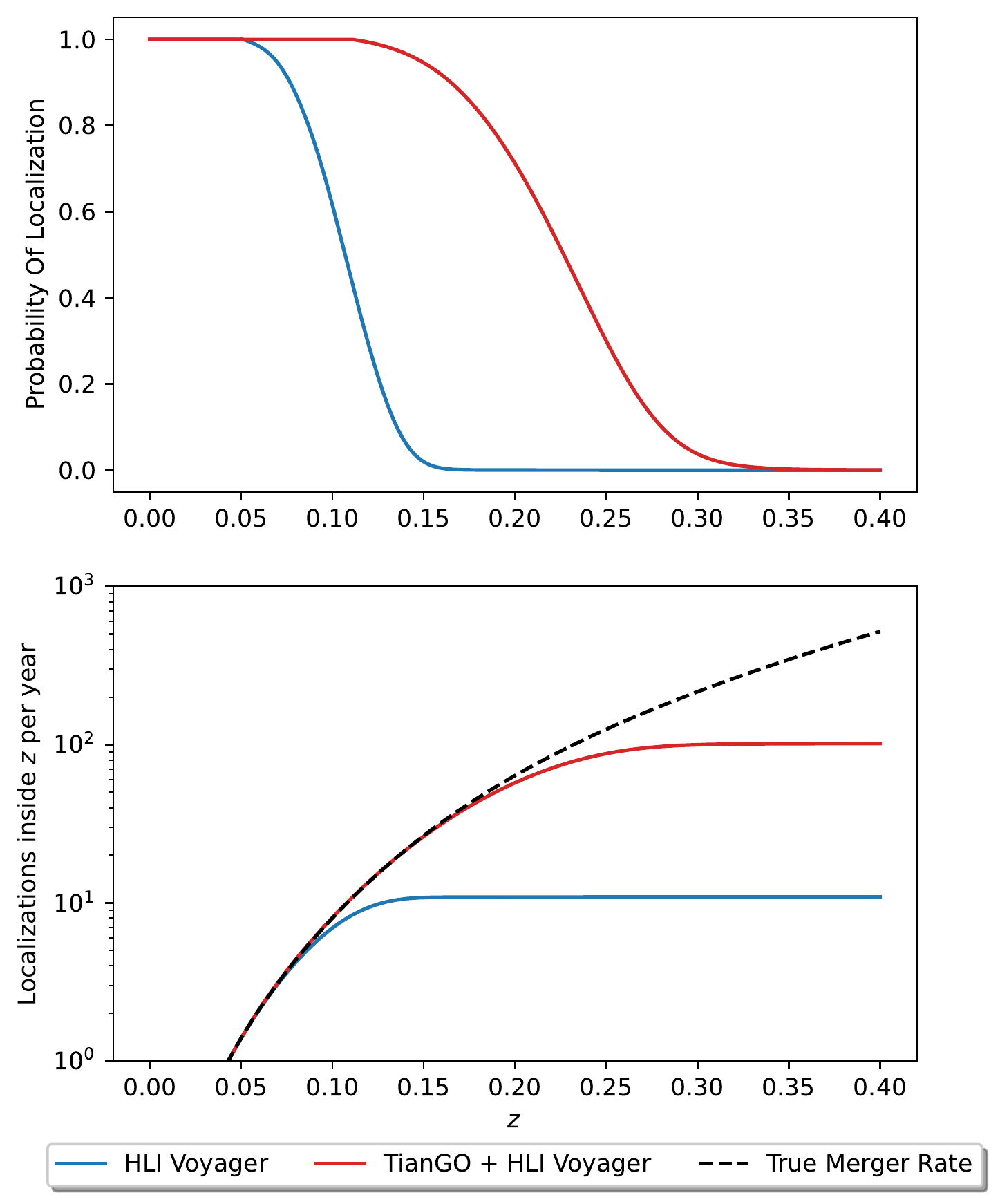}
    \caption{On the top, we plot the probability of an event being localized as a function of redshift for HLI Voyager (green) and TianGO + HLI Voyager (blue). We see that adding TianGO to the HLI Voyager network would nearly double the range at which we can localize a dark siren event.
    On the bottom, we plot the expected number of localizations in the comoving volume sphere. We use the merger rate equal to the star formation rate (red dashed) from Eq.~\eqref{eq:merger-rate}.
    We find that the number of yearly localizations will increase by a factor of 10 by adding TianGO.
    This figure assumes the same binary parameters as Fig.~\ref{fig:PE}, but also uniformly samples the observation time $T_\text{obs} \in [0,5] \text{ yr}$. }
    \label{fig:localization}
\end{figure}

In Fig.~\ref{fig:localization}, we give the number of detections per year which can be fully localized for HLI Voyager with and without TianGO. We see that TianGO will nearly double the range at which a BBH can be localized to a single host. This corresponds to an order of magnitude increase in localization rate. Furthermore, since the localizations occur at higher redshift, we can probe cosmological parameters beyond just the Hubble constant.

%%%%%%%%%%%%%%%%%%%%
%%%%%%%%%%%%%%%%%%%%
%%% Cosmological Constraints %%%
%%%%%%%%%%%%%%%%%%%%
%%%%%%%%%%%%%%%%%%%%

\section{Cosmological Constraints}\label{sec:cosmo-constraints}

\begin{figure}[h]
    \centering
    \includegraphics[width=\linewidth]{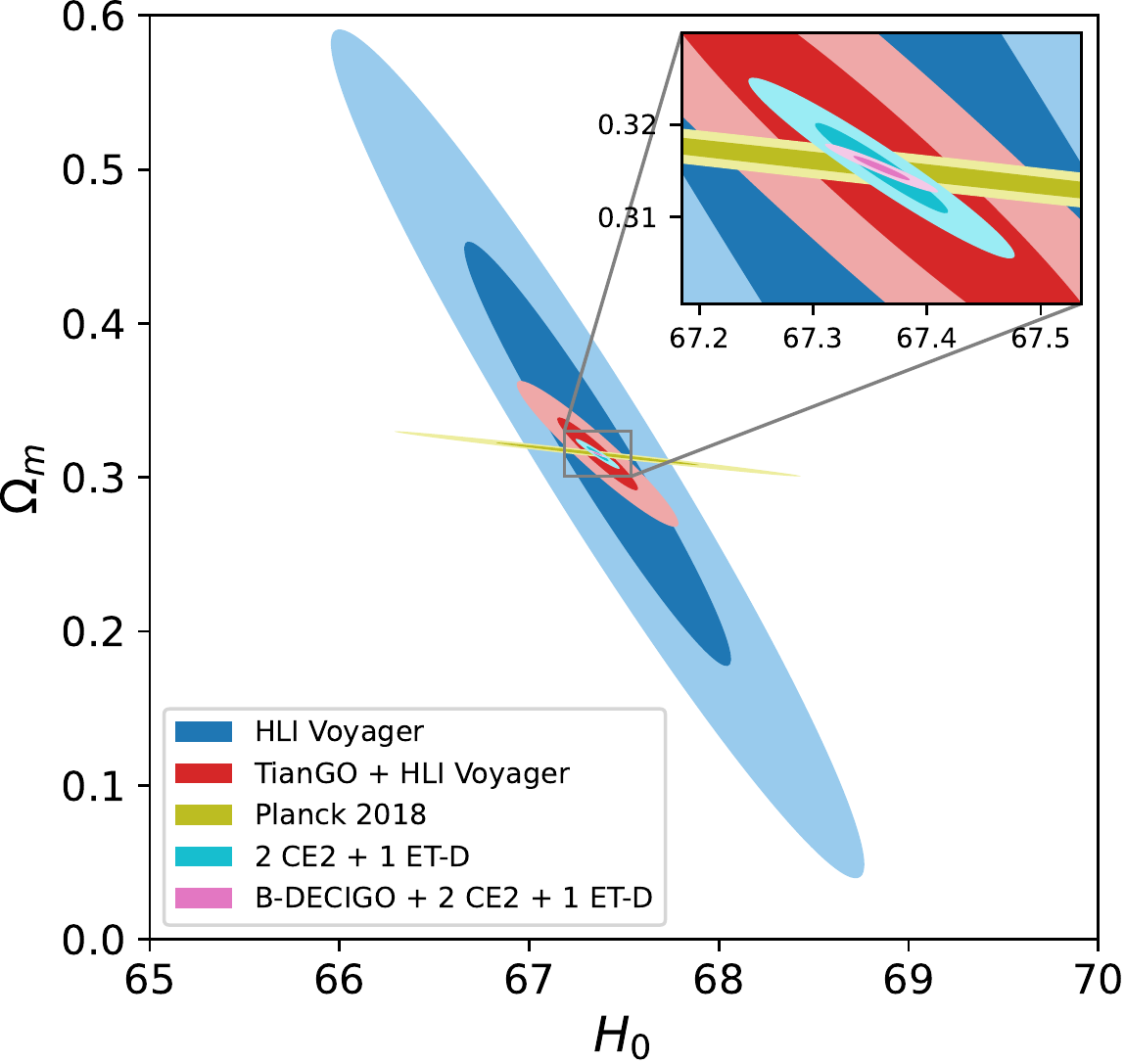}
    \caption{The confidence intervals for Hubble constant $H_0$ and matter density parameter $\Omega_m$ from HLI Voyager (blue), TianGO + HLI Voyager (red), Planck 2018 (yellow) \cite{Planck:2018vyg}, 2 CE 2's + ET-D (cyan), and B-DECIGO + 2 CE 2's + ET-D (pink).
    GW constraints come from Eq.~\eqref{eq:fisher-cosmo} using only fully localized BBH events during a five year observation. We use chirp mass $\mathcal{M}_c = 25 M_\odot$ and merger rate density at the star formation rate. 
    One can see that adding TianGO to the Voyager network would improve error in the measurement of the Hubble constant and the matter density parameter. Moreover, the 3G ground network sees a similar improvement with the addition of B-DECIGO assuming it is in a heliocentric orbit.
    We include the forecasted cosmology constraints for other detector configurations in Tab.~\ref{tab:network-configs}.
    }
    \label{fig:main-plot}
\end{figure}

\begin{figure}[h]
    \centering
    \includegraphics[width=\linewidth]{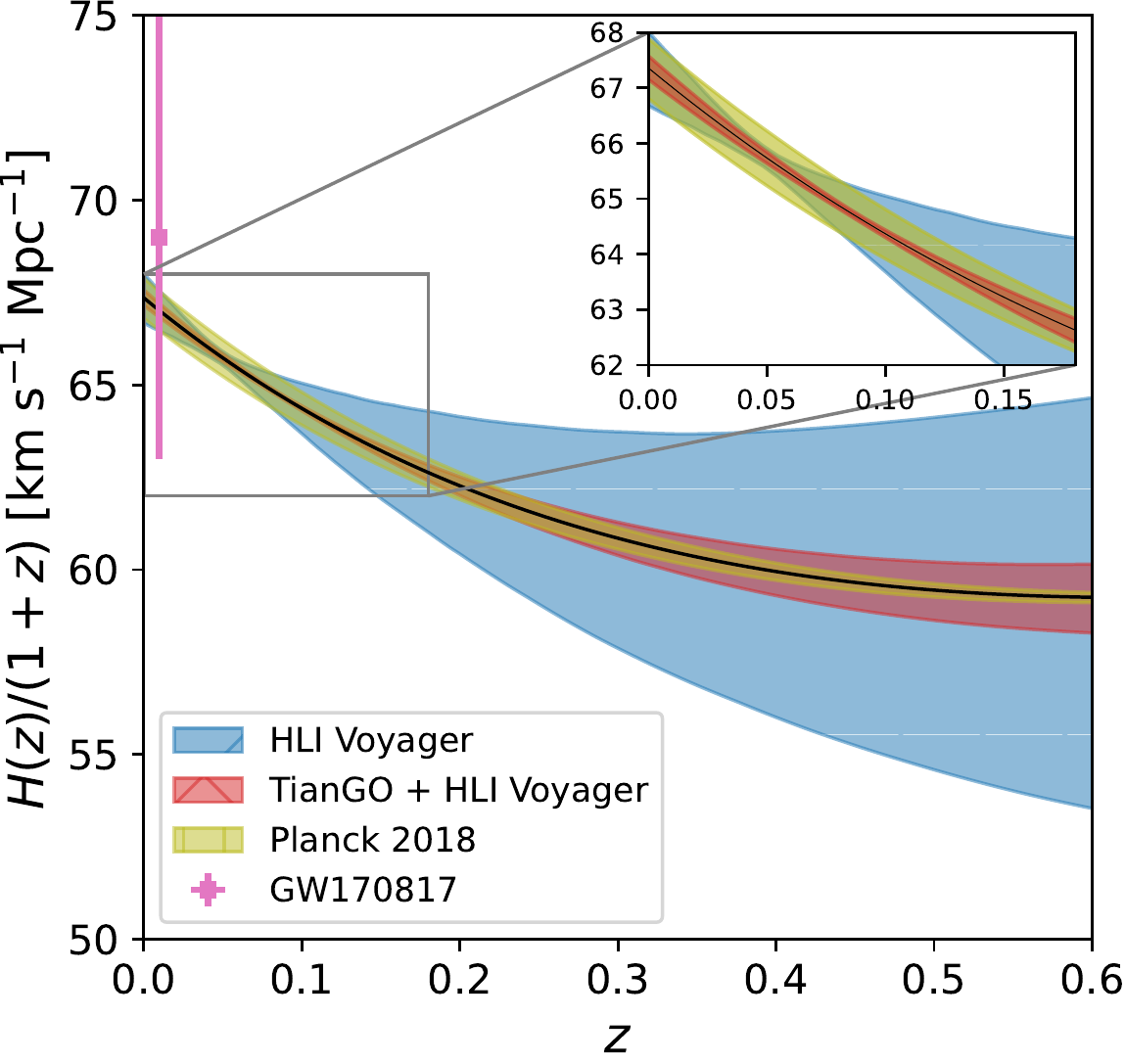}
    \caption{
        Constraints on the expansion rate as a function of redshift for various forecasted and current measurements at the 68\% CL. We plot the forecasted constraints on HLI Voyager (blue) and TianGO + HLI Voyager (red). We also plot current expansion rate constraints from Planck 2018 (yellow) \cite{Planck:2018vyg} and from GW170817 (pink)~\cite{LIGOScientific:2017adf}.
        We produce this plot assuming the Planck parameters as the true values when computing the Fisher matrix, and incorporate only uncertainty on $(H_0, \Omega_m)$ for the shaded regions. Notice that three Voyagers can measure the expansion rate relatively accurately below $z\sim 0.1$. Furthermore, adding the decihertz detector TianGO enhances the ability for the expansion rate to be measured. 
    }
    \label{fig:expansion-rate-plt}
\end{figure}

Given a set of gravitational wave observations, we wish to compute the consistent values of the cosmology. Others have studied how to measure the Hubble constant with dark sirens using statistical inference \cite{Chen:2017rfc, LIGOScientific:2019zcs, Yu:22}. 
Currently, statistical methods are used because the LVK's best localized BBHs have comoving volume resolution of $\Delta V_c \sim 10^5 \text{Mpc}^3$ \cite{LIGOScientific:2021aug} which has thousands of galaxies inside. 
Since our sources are well localized, we can directly measure the redshift of each dark siren event from the uniquely identified host galaxy.
We demonstrate this in 2D with a mock simulation in App.~\ref{app:bayes} that the likelihood function breaks down to the particularly simple answer for well localized sources. We stress that our approach of using the localization condition of $n_\text{gal} \Delta V_C <1$ is a conservative approach. This doesn't require a catalogue since optical telescopes can measure the redshift of the galaxy after the event. Furthermore, galaxy clustering can improve the cosmology constraints \cite{MacLeod:2007jd}. Additionally, more massive galaxies are statistically more likely to be the source of the GW, so this would further improve the ability to localize a GW in the Bayesian approach. Under the localization assumption, a dark siren (BBH) will behave like a bright one (i.e., BNS) for cosmology.

Let us now describe how to compute confidence intervals on the cosmology with a set of dark siren observations. For a set of cosmological parameters $\vect H = (H_0,\Omega_m, ... )$, we can compute their confidence intervals with a Fisher matrix 
\begin{equation}\label{eq:fisher-cosmo}
    \tilde \Gamma_{ij} = \sum_{\text{event } k} \frac{1}{\left(\Delta D_L(z_k)\right)^2} \frac{\partial D_L(z_k,\vect H)}{\partial  H_i}\frac{\partial D_L(z_k,\vect H)}{\partial  H_j} \, ,
\end{equation}
where we use the tilde $\tilde \Gamma$ to distinguish from the waveform parameter estimation matrix used in the last section. Then the error in a cosmological parameter is
\begin{equation}
    \Delta H_i = \sqrt{(\tilde \Gamma^{-1})_{ii}} \, .
\end{equation}
In the nearby universe, the Fisher matrix result reduces to $(\Delta H_0/H_0)^2 = (\Delta D_L/D_L)^2$

In Fig.~\ref{fig:main-plot}, we plot the two sigma confidence intervals on the Hubble constant and matter density parameter using only uniquely localized BBH events. We use a five year observation period, and randomly pick $(\bar\phi_S,\bar\theta_S, \bar\phi_L, \bar\theta_L)$. We use $\mathcal{M}_c = 25 M_\odot$, $q = 1.05$, a trailing angle of $5^\circ$, and uniformly randomize the time until merger. The luminosity distance of the events was sampled accordingly by Eq.~\eqref{eq:merger-rate}. This corresponds to 2515 events with $z<0.4$. There were 43 events localized by HLI Voyager alone and 476 events localized by HLI Voyager + TianGO. 

The addition of TianGO substantially improves our ability to measure the cosmology. Fig.~\ref{fig:main-plot} shows the improvement of using TianGO for measuring the Hubble constant and matter density parameter. Because a multiband measurement increases the distance we can uniquely localize a galaxy, we can measure the matter density parameter much more accurately.
HLI Voyager measures $H_0$ to $1\%$ and $\Omega_m$ to $40\%$, and TianGO upgrades $H_0$ to $0.3\%$ and $\Omega_m$ to $8\%$, while Planck measured $H_0$ to $0.8\%$ and $\Omega_m$ to $2\%$.
We also give the uncertainty ellipse for a possible 3G network consisting of 2 CE2's and 1 ET-D, and also we combine B-DECIGO with the 3G network.
We can see an improvement in both near-term and long-term networks by adding a decihertz detector, particularly in the matter density parameter since its effect is most pronounced at larger redshifts.
Using the covariance matrix containing $(H_0, \Omega_m)$, we can see how well the expansion rate is measured as a function of redshift. In Fig.~\ref{fig:expansion-rate-plt}, we plot the expansion rate $H(z)/(1+z)$ versus redshift where we shade the 68\% CL regions. We can see that gravitational wave detectors are measuring the redshift region $z\sim0.2$ well because the localizations are occurring here because most of localized events are at this redshift. At large redshifts, the cosmic expansion rate uncertainty grows because the matter density parameter is more poorly measured. For reference, we also plot the constraints from GW170817 and Planck 2018. Note that we only show Fig.~\ref{fig:expansion-rate-plt} up to $z = 0.6$ since we cannot measure $\Omega_\Lambda, \Omega_k$ well enough with localized BBH sources.

Finally, we estimate the constraints on the Hubble constant and matter density parameter for various 2G to 3G detector networks in Tab.~\ref{tab:network-configs}. 
Specifically, we compare the cosmological constraints from localized dark sirens during a 5 year observation period.
For the 3G detectors, we consider Cosmic Explorer 2 (CE2), and Einstein Telescope D (ET-D).
We see that even with 2 CE2's and ET-D, TianGO improves the ability to measure the Hubble constant by a factor of 2, and the matter density parameter by a factor of 3. This is because we see a sizable improvement in the number of localized events.

For the long-term multiband case, we use a network consisting of B-DECIGO, CE2, and ET-D. Because the orbit of B-DECIGO is still under discussion \cite{Kawamura:2018esd}, we placed it in a trailing 5$^\circ$ orbit like TianGO. We performed the same analysis as in Section~\ref{sec:cosmo-constraints}. We find that the addition of B-DECIGO can improve the cosmological measurement capabilities of the 3G detectors.

% \usepackage{graphicx}
\begin{table*}[]
\centering
\resizebox{\textwidth}{!}{%
\begin{tabular}{|l|l|l|l|l|}
\hline
                                                                      & $\Delta H_0 / H_0$                                                                 & $\Delta \Omega_m$                                                                  & Localizations / 5 yr                                 & Notes                                                                                                                                                 \\ \hline
\begin{tabular}[c]{@{}l@{}}3 V \\ (+ T)\end{tabular}                  & \begin{tabular}[c]{@{}l@{}}$1 \times 10^{-2}$ \\ ($2 \times 10^{-3}$)\end{tabular} & \begin{tabular}[c]{@{}l@{}}$1 \times 10^{-1}$\\ ($2  \times 10^{-2}$)\end{tabular} & \begin{tabular}[c]{@{}l@{}}43\\ (476)\end{tabular}   & \begin{tabular}[c]{@{}l@{}}Voyager at Hanford, Livingston, India\\ sites.\end{tabular}                                                                \\ \hline
\begin{tabular}[c]{@{}l@{}}1 CE2 + 1 ET-D\\ (+ T)\end{tabular}        & \begin{tabular}[c]{@{}l@{}}$2 \times 10^{-3}$\\ ($6 \times 10^{-4}$)\end{tabular}  & \begin{tabular}[c]{@{}l@{}}$1 \times 10^{-2}$\\ ($3 \times 10^{-3}$)\end{tabular}  & \begin{tabular}[c]{@{}l@{}}382\\ (1930)\end{tabular} & \begin{tabular}[c]{@{}l@{}}CE2 at Hanford, \\ ET-D at GEO-600 sites.\end{tabular}                                                                     \\ \hline
\begin{tabular}[c]{@{}l@{}}2 CE2 + 1 ET-D\\ (+ T)\end{tabular}        & \begin{tabular}[c]{@{}l@{}}$1 \times 10^{-3}$\\ ($5 \times 10^{-4}$)\end{tabular}  & \begin{tabular}[c]{@{}l@{}}$6 \times 10^{-3}$\\ ($2 \times 10^{-3}$)\end{tabular}  & \begin{tabular}[c]{@{}l@{}}843\\ (2410)\end{tabular} & \begin{tabular}[c]{@{}l@{}}CE2 at Hanford, Livingston.\\ ET-D at GEO-600 sites.\end{tabular}                                                          \\ \hline
\begin{tabular}[c]{@{}l@{}}2 CE2 + 2V\\ (+ T)\end{tabular}            & \begin{tabular}[c]{@{}l@{}}$1 \times 10^{-3}$\\ ($6 \times 10^{-4}$)\end{tabular}  & \begin{tabular}[c]{@{}l@{}}$9 \times 10^{-3}$\\ ($3 \times 10^{-3}$)\end{tabular}  & \begin{tabular}[c]{@{}l@{}}556\\ (2211)\end{tabular} & \begin{tabular}[c]{@{}l@{}}CE2 at Virgo, India sites. Voyager at \\ Hanford, Livingston sites.\end{tabular}                                           \\ \hline
\begin{tabular}[c]{@{}l@{}}1 CE2 + 1 ET-D\\ (+ B-Decigo)\end{tabular} & \begin{tabular}[c]{@{}l@{}}$2 \times 10^{-3}$\\ ($5 \times 10^{-4}$)\end{tabular}  & \begin{tabular}[c]{@{}l@{}}$1 \times 10^{-2}$\\ ($2 \times 10^{-3}$)\end{tabular}  & \begin{tabular}[c]{@{}l@{}}380\\ (4758)\end{tabular} & \begin{tabular}[c]{@{}l@{}}CE2 at Hanford, ET-D at GEO-600 sites.\\ B-Decigo placed in 5$^\circ$ trailing Heliocentric orbit\end{tabular}             \\ \hline
\begin{tabular}[c]{@{}l@{}}2 CE2 + 1 ET-D\\ (+ B-Decigo)\end{tabular} & \begin{tabular}[c]{@{}l@{}}$1 \times 10^{-3}$\\ ($3 \times 10^{-4}$)\end{tabular}  & \begin{tabular}[c]{@{}l@{}}$6 \times 10^{-3}$\\ ($1 \times 10^{-3}$)\end{tabular}  & \begin{tabular}[c]{@{}l@{}}835\\ (5770)\end{tabular} & \begin{tabular}[c]{@{}l@{}}CE2 at Hanford, Livingston, ET-D at GEO-600 sites.\\ B-Decigo placed in 5$^\circ$ trailing Heliocentric orbit\end{tabular} \\ \hline
\end{tabular}%
}
\caption{Dark siren constraints on the Hubble constant and matter density parameter for various detector configurations. We use the same methodology as for this table as in the rest of this paper. We find the Fisher matrix confidence interval on the cosmological parameters by using only dark sirens which are completely localized.
}
\label{tab:network-configs}
\end{table*}

\section{Conclusion}\label{sec:conclusion}

In this paper, we studied how a space-based decihertz detector can enhance the sensitivity of a ground network for dark siren cosmological measurement. 
We construct the case that these detectors will measure a significant number of `bright' dark siren BBH -- GW from which we can uniquely localize and uniquely identify the host galaxy.
We then use a Fisher matrix formalism to place constraints on the cosmological parameters.
We estimated how well the Hubble constant and matter density parameter could be measured by BBH dark sirens with a five year observation of TianGO plus three LIGO Voyagers. 
The result is the multiband detection of dark sirens improves the measurement of the Hubble constant by about a factor of 3. The larger redshift localized events allows the matter density parameter to be resolved in the multiband case.

In the future, it would be interesting to extend our analysis to include dark sirens which are non-uniquely identified, but are still well localized. 
Since the fully localized criterion leaves out events with just a small number of galaxies, information about the cosmology can still be extracted from these events. 
Moreover, there are other effects which can improve the sensitivity further, such as exploiting the clustering of galaxies to improve localization \cite{MacLeod:2007jd} and weighting the galaxies by luminosity \cite{Gray:2019ksv}.

Measuring the cosmology with gravitational waves is easier when the host galaxy is uniquely identified. 
The statistical dark siren approach is degenerate with parameters such the merger rate evolution with redshift and the BBH population model (as discussed in the GWTC-3 cosmology paper \cite{LIGOScientific:2021aug}). 
Simultaneously measuring the cosmology and these population parameters can be done by looking at the distribution of BBH events \cite{You:21, Mukherjee:21, Yu:22}, but would result in a less sensitive measurement of the cosmological parameters.
Otherwise, if these factors are not jointly measured, this would bias the measurement of the Hubble constant \cite{Trott:2021fnx, Yu:22}. 
Consequently, a multiband detection of dark sirens with uniquely identified hosts has the potential to isolate the measurement of cosmological parameters from these population parameters.

\begin{acknowledgements}
B.S. acknowledges support by the National Science Foundation Graduate Research Fellowship under Grant No. DGE-1745301.
H.Y. acknowledges the support of the Sherman Fairchild Foundation. Y.C. and B.S. acknowledge support from the Brinson Foundation, the Simons Foundation (Award Number 568762), and by NSF Grants PHY-2011961, PHY-2011968, PHY--1836809.
\end{acknowledgements}

\appendix
\onecolumngrid 
\section{Antenna Patterns of TianGO}\label{app:antenna}

The standard formula for the plus and cross antenna patterns of a detector is
\begin{align}
    F_+ &= \left( \frac{1+\cos^2\theta_S}{2} \right) \cos 2\phi_S \cos 2\psi_S-\cos\theta_S \sin2\phi_S \sin2\psi_S \, ,\\
    F_\times &= \left( \frac{1+\cos^2\theta_S}{2} \right) \cos 2\phi_S \cos 2\psi_S + \cos\theta_S \sin2\phi_S \sin2\psi_S \, .
\end{align}
where $(\phi_S,\theta_S)$ are in the detector frame. We use the pycbc detector class to get the ground based antenna patterns \cite{Biwer:2018osg}. The antenna patterns of a space detector are more complicated however, because the detector has changing orientation. This means that the antenna patterns have time dependence $F_{+,\times}(t)$, which we will use the time frequency relation to find their frequency dependence.

To find the detector beam pattern coefficients, let us first describe the geometry of the system. We have two coordinate systems: unbarred coordinates $(\uvect{x},\uvect{y},\uvect{z})$ which correspond to the individual detector and barred coordinates $(\uvect{\bar{x}},\uvect{\bar{y}},\uvect{\bar{z}})$ in the ecliptic frame. The relationship between the orientation of the detector frame and the ecliptic is
\begin{align}
    \uvect{x}(t) &=-\frac{\sin 2 \phi_{\mathrm{t}}}{4} \uvect{\bar{x}}+\frac{3+\cos 2 \bar{\phi}_{\mathrm{t}}}{4} \uvect{\bar{y}}+\frac{\sqrt{3}}{2} \sin \bar{\phi}_{\mathrm{t}} \uvect{\bar{z}} \, ,\nn\\
    \uvect{ y}(t) &= \uvect z(t) \times \uvect x(t)\, , \nn\\
    \uvect{z}(t)&=-\frac{\sqrt{3}}{2}\left(\cos \bar{\phi}_{\mathrm{t}} \uvect{\bar{x}}+\sin \bar{\phi}_{\mathrm{t}} \uvect{\bar{y}}\right)+\frac{1}{2} \uvect{\bar{z}} \,, \label{eq:detector-coords}
\end{align}
where the phase of TianGO in the ecliptic frame is equal to 
\begin{equation}\label{eq:phit}
    \bar{\phi}_t(f) = \frac{2 \pi t(f)}{1 \text{ yr}} - t_a \, ,
\end{equation}
where $t_a$ is the trailing angle, and equal to $5^\circ$ for TianGO. The time as a function of frequency is \cite{Kuns:2019upi} 
\begin{equation}
    t(f) = t_c - 5 \left( 8 \pi f \right)^{-8/3} \mathcal{M}_z^{-5/3} \left[ 1 + \frac{4}{3} \left( \frac{743}{336} + \frac{\mu}{M} x - \frac{32 \pi}{5}x^{3/2}\right) \right] \, ,
\end{equation}
where $\mu$ is the reduced mass and 
\begin{equation}
    x = \left( \pi M_z f \right)^{2/3} \, .
\end{equation}

We can now write $(\phi_S(f),\theta_S(f),\psi_S(f))$ for the TianGO detector using Eq.~\eqref{eq:detector-coords},
\begin{align}
    \cos \theta_S(f) &= \frac{1}{2} \cos \bar\theta_S - \frac{\sqrt{3}}{2} \sin\bar\theta_S \cos\left( \bar\phi_t(f) - \bar\phi_S \right) \, , \\
    \phi_S(f) &= \bar\phi_t(f) + \arctan \left[ \frac{\sqrt{3}\cos\bar\theta_S + \sin\bar\theta_S \cos \left( \bar\phi_t(f)-\bar\phi_S \right)}{2 \sin\bar\theta_S \sin\left( \bar\phi_t(f)-\bar\phi_S \right)} \right] \, .
\end{align}
The polarization phase of TianGO is 
\begin{equation}
    \tan \psi_S(f) = \frac{\uvect L \cdot \uvect z - \left( \uvect L \cdot \uvect N \right)\left( \uvect z \cdot \uvect N \right)}{\uvect N \cdot \left( \uvect L \times \uvect z \right)}
\end{equation}
where
\begin{align}
    \uvect N \cdot \uvect z &= \cos\theta_S(f) \, , \\
    \uvect L \cdot \uvect z &= \frac{1}{2} \cos \bar\theta_L - \frac{\sqrt{3}}{2} \sin \bar\theta_L \cos \left[ \bar\phi_t(f) - \bar\phi_L \right] \, ,\\
    \uvect L \cdot \uvect N &= \cos\bar\theta_L \cos\bar\phi_S + \sin\bar\theta_L \sin\bar\theta_S \cos \left( \bar\phi_L -\bar\phi_S \right) \, , \label{eq:LdN}\\
    \uvect N \cdot \left( \uvect L \times \uvect z \right) &= \frac{1}{2} \sin \bar{\theta}_{L} \sin \bar{\theta}_{S} \sin \left(\bar{\phi}_{L}-\bar{\phi}_{S}\right) \nn \\
    -\frac{\sqrt{3}}{2} &\cos \bar{\phi}_t(f)\left(\cos \bar{\theta}_{L} \sin \bar{\theta}_{S} \sin \bar{\phi}_{S}-\cos \bar{\theta}_{S} \sin \bar{\theta}_{L} \sin \bar{\phi}_{L}\right) \nn\\
    -\frac{\sqrt{3}}{2} &\sin\bar{\phi}_t(f)\left(\cos \bar{\theta}_{S} \sin \bar{\theta}_{L} \cos \bar{\phi}_{L}-\cos \bar{\theta}_{L} \sin \bar{\theta}_{S} \cos \bar{\phi}_{S}\right)\, .
\end{align}

\section{Consistency of Statistical Method}\label{app:bayes}

In the statistical method, we wish to break the $z - D_L$ degeneracy by using a galaxy catalog with the gravitational wave observation. We will use the method described in a variety of sources \cite{Chen:2017rfc,Gray:2019ksv}. If we wish to constrain the cosmological parameters $\vect H$ and have gravitational wave data $d_\text{GW}$, then with Bayes theorem, we have
\begin{equation}
    p(\vect H | \dgw) \propto p(H_0) p(\dgw | \vect H)
\end{equation}
where
\begin{align}\label{eq:bayes_like}
    p(\dgw | \vect H) &= \frac{1}{\beta(\vect H)}\int p(\dgw, D_L, \phi_S,\theta_S, z| \vect H) d D_L d \phi_S d\theta_S dz \, , \\
    &= \frac{1}{\beta(\vect H)} \int p(\dgw| D_L(z, \vect H), \phi_S, \theta_S) p_0(z,\phi_S,\theta_S) d \phi_S d\theta_S dz \, .\label{eq:bayes_like2}
\end{align}
the first term in the integral is approximated with a multivariate Gaussian distribution
\begin{equation}
    p(d_\text{GW}| D_L(z, \vect H), \phi_S, \theta_S) = N(D_L(z,\vect H) - \hat D_L,\sigma_{D_L}^2) N(\phi_S - \hat\phi_S,\sigma_{\phi_S}^2)N(\theta_S - \hat\theta_S,\sigma_{\theta_S}^2) \, ,
\end{equation}
where $N(x - \mu,\sigma^2)$ is the probability density function of the normal distribution, $(\hat D_L, \hat \phi_S, \hat \theta_S)$ are the true event parameters, and $(\sigma_{D_L}, \sigma_{\phi_S}, \sigma_{\theta_S})$ is given by the Fisher matrix analysis in Eq. (10). The second term in the integral is the galaxy catalog
\begin{equation}\label{eq:catalogue-sum}
    p_0(z,\phi_S,\theta_S|\vect H) = \frac{1}{N_\text{gal}} \sum^{N_\text{gal}}_i N(z- z^i,\sigma_{z_i}^2)\delta(\phi_S- \phi_S^i)\delta(\theta_S - \theta_S^i) \, ,
\end{equation}
where $\sigma_{z_i}$ is the variance due to the peculiar velocity. The variables $(z^i, \phi^i,\theta^i)$ are the mean redshift and angular location of the $i$th galaxy, while unbarred variables are parameters.

The angular uncertainty is negligible and the distribution is replaced with a Dirac delta function $\delta(\phi_S- \bar\phi_S^i)$ and similarly for $\theta_S$. Finally, the normalization $\beta(\vect H)$ is 
\begin{equation}
    \beta(\vect H) = \int_{d_{\rm GW}> d_{\rm GW}^{\rm th}} p(\dgw, D_L, \phi_S,\theta_S, z| \vect H) d D_L d \phi_S d\theta_S dz  \,d \dgw \, ,
\end{equation}
where 
\begin{equation}
    p(\dgw, D_L, \phi_S,\theta_S, z| \vect H) =  p(\dgw| D_L(z, \vect H), \phi_S, \theta_S) p_0(z,\phi_S,\theta_S) \, .
\end{equation}
and where $d_{\rm GW}^{\rm th}$ is the detection threshold.
Note that Eq.~\eqref{eq:bayes_like} reduces to the Fisher matrix confidence interval Eq.~\eqref{eq:fisher-cosmo} on $\vect H$ if only one galaxy has nonvanishing likelihood. This reduction can be derived by examining \eqref{eq:bayes_like2} in the case that there is only one galaxy inside the volume. This happens when all other galaxies in the sum in $p_0(z,\phi_S,\theta_S|\vect H)$ do not contribute to the integral in Eq.~\eqref{eq:bayes_like2}.

Now, let us demonstrate the statistical method in 2D and examine its convergence as a function of the number of galaxies inside the localization region. We assume that $D_L = z/H_0$ and that the peculiar velocity uncertainty is subdominant. Thus, we assume the peculiar velocity is a very sharp Gaussian and absorb it into $\sigma_{D_L}$. If we call $h = (H_0) / (H_0)_\text{true}$, the likelihood function is
\begin{equation} \label{eq:single-plt}
    p(\dgw | h ) = \frac{1}{\beta(h)}\frac{1}{N_\text{gal}} \sum_{i} N(\hat D_L-D_L^i(h), \sigma_{D_L}^2) N(\hat \phi_S - \phi_S^i,\sigma_{\phi_S}^2)N(\hat \theta_S - \theta_S^i,\sigma_{\theta_S}^2)
\end{equation}
where $D_L^i(h)=z^i/H_0=z^i/[h(H_0)_{\rm true}]$ and $\sigma_{z} = \sigma_{D_L}  (H_0)_\text{true}$. In this 2D case, $\beta(h) \propto h^2$.

\begin{figure}[h]
    \centering
    \includegraphics[width=0.4\textwidth]{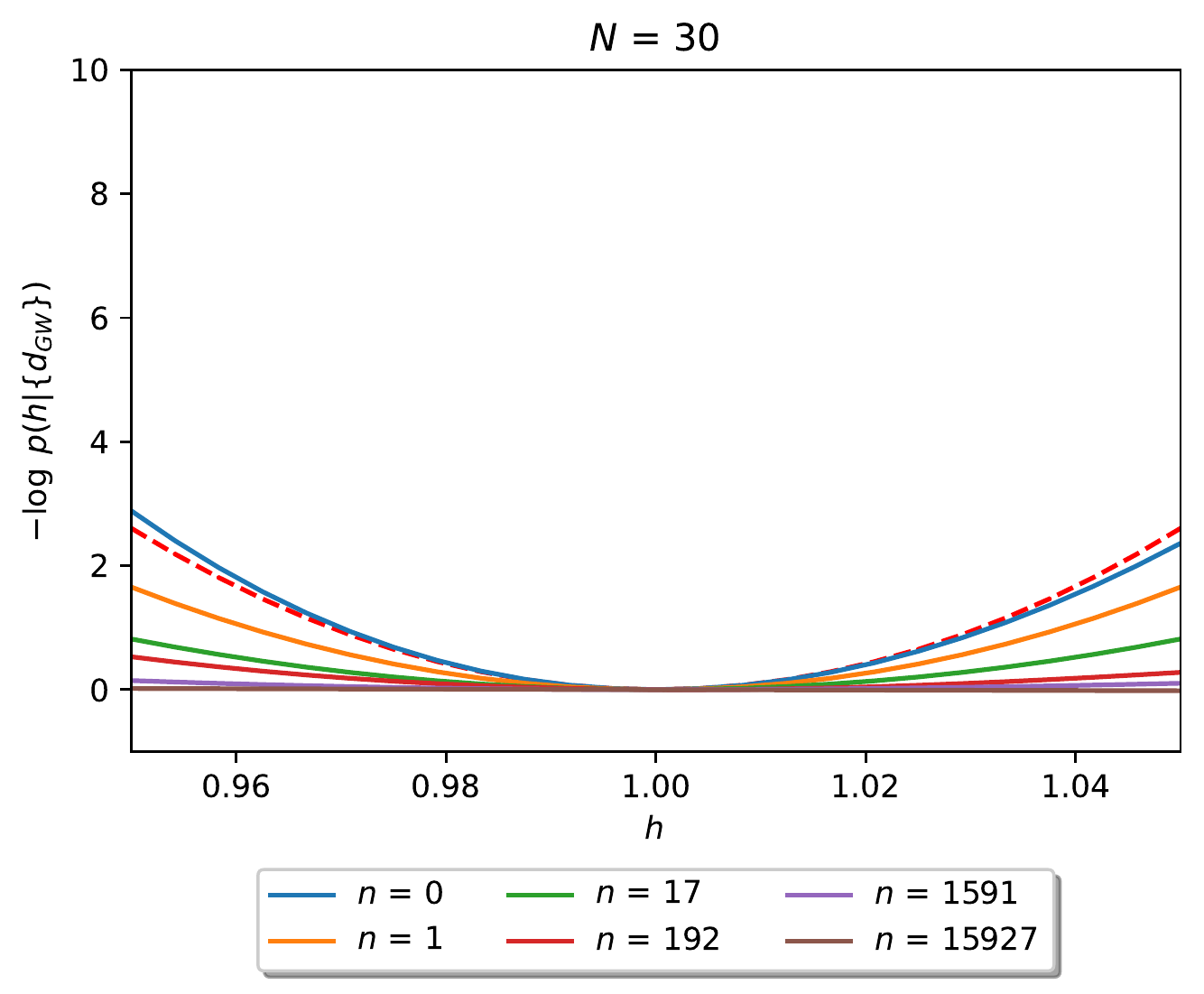}
    \includegraphics[width=0.4\textwidth]{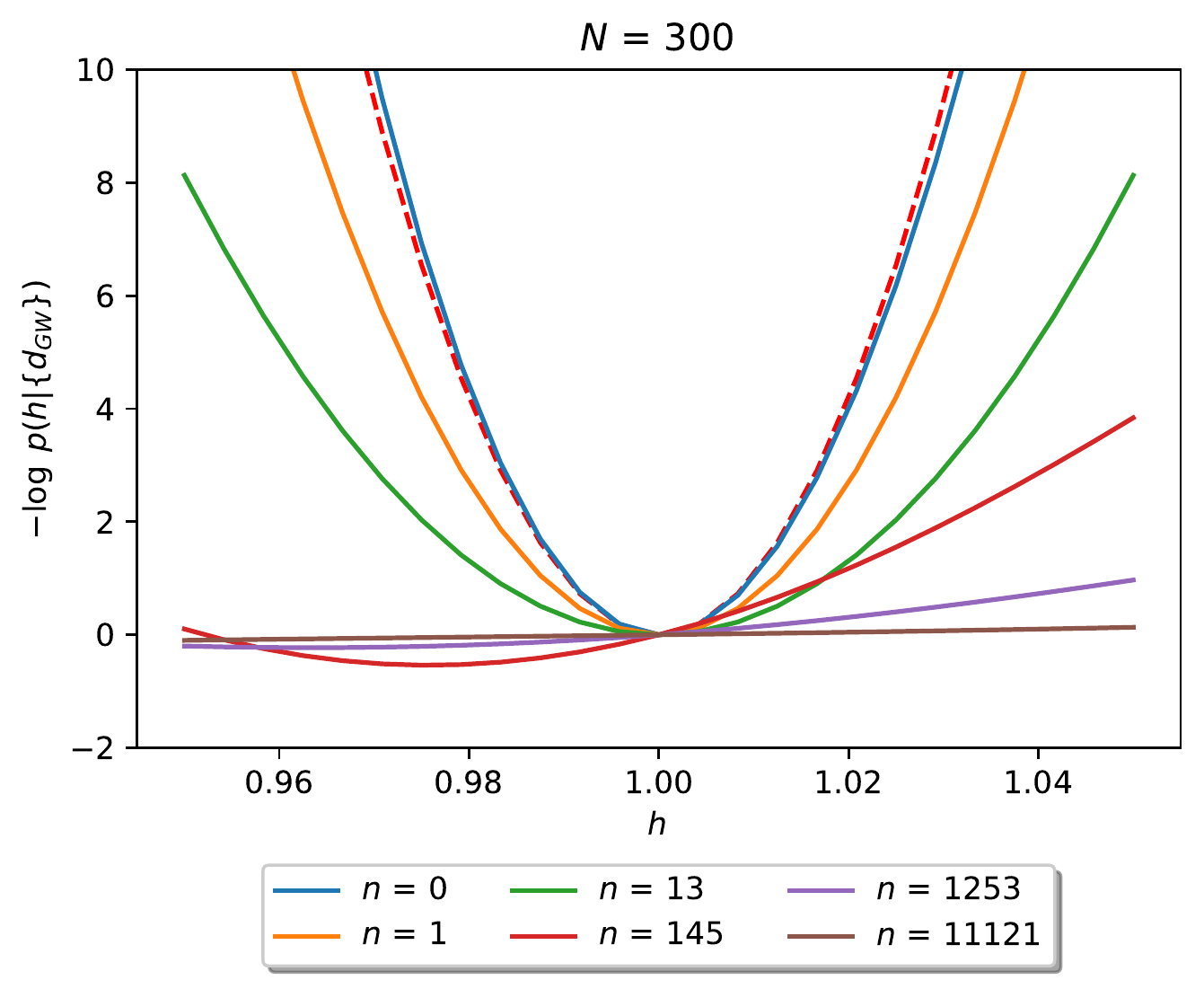}
    \caption{
        We plot the negative log likelihood of the posterior distribution on the Hubble constant $p(h | \left\{ \dgw \right\})$ in the 2D simulation.
        Each curve with $n$ labeled is the median number of \textit{extra} galaxies in the localization region while the Fisher matrix constraint approximation from Eq.~\eqref{eq:fisher-cosmo} is also plotted (dashed red).
        Each curve in the plot corresponds to picking a different angular resolution for the events. This shows that measurement of sources with poor angular resolution will result in weaker Hubble constant constraints due to the increased number of galaxies in the localization region.
        We also see that with a higher number of events, the likelihood distribution for $h$ tightens.
        Finally, we see that the Bayesian approach reduces to the Fisher information estimate when there is a uniquely identified galaxy. This is still a conservative estimation on how well we can measure the cosmology as the information from $n>0$ systems is discarded.
        }
    \label{fig:LL-vs-ngal}
\end{figure}

If we need to stack events, we generalize Eq.~\eqref{eq:single-plt} to be the product of the likelihood function of each event\footnote{Technically, there is another factor $p(N | h)$ in front of the product which depends on the intrinsic astrophysical merger rate and comoving volume surveyed. It is discussed after Eq.~(7) in Ref.~\cite{Gray:2019ksv}.},
\begin{equation}
    p(\left\{ \dgw \right\} | h ) = \prod_{\text{event } e}^N p( \left( \dgw \right)_e | h ) \, .
\end{equation}
If we assume a uniform prior on $h$, then $p(h | \left\{ \dgw \right\}  ) \propto p(\left\{ \dgw \right\} | h )$.  In Fig.~\ref{fig:LL-vs-ngal}, we plot the posterior on $h$ for 30 and 300 events. In this figure, we vary the angular resolution of the events for each curve. We plot the median number of potential host galaxies for the events. One can see that as events are nearly perfectly localized $(n \rightarrow 0)$, the posterior on $h$ approaches the Fisher likelihood in Eq.~(\ref{eq:fisher-cosmo}).

Due to the potential systematics possible in such an experiment, we list the precise choices we used to make the plot. Our distance resolution was $\Delta D_L / D_L = 0.15 z + 10^{-2}$ and our angular resolutions varied between $\Delta \phi_S = \frac{z}{1000} \deg$ to $\Delta \phi_S = 100 z \deg$. These scaled with redshift linearly due to the SNR scaling of parameter measurement, while the $10^{-2}$ is the same order as the peculiar velocity error (so a few close events don't dominate). We uniformly placed $3\times 10^6$ galaxies thoughout the disc in the $z\in[0,2)$ 'redshift window'. For each event, we randomly picked a galaxy with $z\in [0,1)$. The particular redshift window can have  a systematic effect on the statistical method \cite{Trott:2021fnx}, and we chose our galaxy disc to be much bigger than the redshift window to avoid artificial boundary effects.

\bibliography{bibliography.bib}

\end{document}